\pdfoutput=1
\documentclass[10pt]{article}

\usepackage{color}
\usepackage{bbm}
\usepackage{epsfig}
\usepackage{amssymb,amsmath,amscd, mdwmath}
\usepackage{graphicx,cite,url}
\usepackage{algorithmic}
\usepackage{float}
\usepackage{amsfonts}
\usepackage{syntonly}
\usepackage{multirow}
\usepackage{graphicx}
\usepackage{subfig}
\usepackage{tabularx}
\usepackage{setspace}
\usepackage{stfloats}
\usepackage{amsthm}
\usepackage[labelfont=bf,labelsep=period,justification=raggedright]{caption}

\makeatletter
\renewcommand{\@biblabel}[1]{\quad#1.}
\makeatother

\date{}

\begin{document}

\begin{flushleft}
{\Large
\textbf{Profiling MOOCs    from viewing perspective}
}
\\
Zheng Xie$^{1, \sharp }$
Xiao Xiao$^{2}$
Jianping Li$^{1}$
Jinying Su$^{1}$
\\
\bf{1}  College of Arts and Sciences,   National University of Defense Technology, Changsha,   China\\
\bf{2} iCourse, Higher Education Press, Beijing,   China\\
$^\sharp$ xiezheng81@nudt.edu.cn\\
 The   authors have  contributed equally to this work.
 \end{flushleft}
\section*{Abstract}

We profiled three aspects of MOOCs from the perspective of viewing behaviors, the most prominent and common ones of MOOC learning. They were learner classification, course attraction, teaching order and learning order. Based on viewing behavior data, we provided a non-parametric algorithm to categorize learners, which helped to narrow the scope of finding potential all-rounders, and a method to measure the correlations between teaching order and learning order, which helped to assign teaching contents. Using information entropy, we provided an index to measure course attraction, which integrated the viewing time invested on courses and the number of viewed course videos. This index describes the diminishing marginal utility of repeated viewing and the increasing information of viewing new videos. It has potential to be an auxiliary method of assessing course achievements.



\section*{Introduction}

Massive open online courses (MOOCs) have emerged from the integration of education and the Internet\cite{Breslow}. They break the boundaries of time and space, expanding traditional education due to their transmission of information by the Internet technology. And they have been viewed as an accelerator for learning and a solution to educational resource imbalance\cite{Reich,Emanuel}. Differences between traditional courses and MOOCs lie in several dimensions, involving conditions of admission  (pretesting vs. no-condition), student motivations (homogeneous vs. heterogeneous), classroom management (supervised vs. unsupervised), interactions (face-to-face vs. online), dropout rates (low vs. high)\cite{Murphy,Anderson}.
Moreover,
 MOOCs are featured as  learner-centered, which is different from the knowledge-centered feature of  traditional education.
 Therefore, understanding    MOOC learning behaviors helps to assess  MOOC achievements,
  to   find  methods of improving MOOC quality, and so on.


Analyzing
 MOOC learning behaviors
 has become a hot topic in the  MOOC   community, which includes learning motivations, learning achievements,  and so on\cite{Hew,Meyer,DeBoer,Jona2014}.
 MOOC learners are motivated not just to pass exams which involve understanding particular concepts, or some parts of course contents\cite{Barba,Watted,Zheng,Barak}.
Their diversified expectations and motivations to learn MOOCs result in high course dropout rates and low exam participation rates\cite{Hone, Onah,Freitas,Greene,Liyanagunawardena}. Viewing behaviors are the most prominent and common in MOOC learning, compared with other behaviors such as doing exercises, discussing and testing.
Therefore,
profiling MOOCs    from viewing perspective can involve as many learners as possible.

We profiled MOOCs from viewing perspective in following three aspects. Firstly, we provided a method to categorize learners into two types,  which helped to narrow the  search range of potential all-rounders.
 Secondly, we provided an index to measure course attraction based on course learners' the number of viewed videos (calculated in a continuous way) and their relative viewing time length compared with video length. Thirdly, we provided a method to measure the correlations between teaching order and learning order based on learners' viewing order and video labels, which helped to optimize teaching content assignment.

MOOC learners' behavior data cannot inherently pose  answers to assess courses because the causal relationship between learning more and learning better is unclear\cite{Henrie}. So our results might not  be the exact   MOOC contributions to learners. However, the low values  of those indexes can help  us to find some imperfect aspects of some MOOCs. Note that the order correlation cannot be applied to humanities courses, but can     to natural science courses.











This paper is organized as follows. The    data are described in Section  2.  The indexes such as  entropy    are described
   in Section 3.  The    indexes of attractions and those of  the correlations between teaching order and learning order  are described in Sections 4 and 5. The conclusion is drawn in Section 6.

\section*{Viewing behavior data}

MOOC platform iCourse (http://www.icourse163.org) provided the viewing behavior data of eight courses (01/01/2017--10/11/2017). The courses were selected from natural sciences, social sciences, humanities and engineering technology. Each course had substantial registrants so that our results were statistically meaningful. The data included time length of each video. For each learner, the data included
 the  viewing start time
 and the viewing time length of  each video he viewed.

Since some selected courses were not finished before 10/11/2017, our discussions focused on the measurements of course attractions on the level of videos, and on the measurements of correlations between teaching order and viewing order, to which the data of some weeks were adequate. Videos could only be downloaded by iCourse app. If the app disconnected to the Internet, the information of viewing downloaded videos cannot be collected. Accordingly, our discussions only involved online viewing behaviors of MOOCs.

\begin{table*}[!ht] \centering \caption{{\bf Specific statistical indexes of   the  data provided by{ ICourse}. } }
\footnotesize\begin{tabular}{l rrrrrr rr } \hline
 Course  & Course Id& $a$ &	$b$&	$c$ & $d$ & $e$& $f$\\ \hline
{\it Calculus}&1002301004&	2,955&	129	&8.081&	0.998&	0.189&	2\\
{\it Game theory}&	1002223009&4,764&	38	&7.141	&2.238&	0.427	&66\\
{\it Finance	}& 1002301014  &6,380	&63	&5.368&	1.310&	0.330	&2\\
{\it Psychology}&	1002301008&3,827	&26	&5.008	&0.913&	0.204	&59\\
{\it Spoken English}& 1002299019	&11,719&	46	&3.032&	0.321&	0.106&	7\\
{\it Etiquette}&1002242007	&3,846&	41	&7.787	&1.271&	0.205&	22\\
{\it C Language}&1002303013 &	17,541&	81	&12.47&	1.541&	0.142&	39\\
{\it Python}&	1002235009&13,417&	53	&10.32&	0.896&	0.087&	28\\
\hline
 \end{tabular}
  \begin{flushleft}
    Index  $a$: the number of learners, $b$: the number of videos,
  $c$: the number of videos viewed by per learner,
  $d$:   the viewing time length	per learner  (unit: hour), $e$: the time length	per video (unit: hour), and $f$: the number of all-rounders.
\end{flushleft}
\label{tab1}
\end{table*}

   Specific  statistical indexes of   viewing behaviors were listed in Table~\ref{tab1}, which
    can be used to measure the influence of video lengths on  completion rates  of viewing
videos.
     Suppose   learners $\{L_1,...,L_m\}$ view   a course  with $n$ videos $\{V_1,...,V_n\}$.
For each leaner $L_s$ ($s=1,...,m$), denote the label set of he viewed videos as $S^V_s$.
For each video $V_i$ ($i=1,...,n$),   denote the label set of learners who viewed it as $S^L_i$.
 Denote the time length of video $V_i$   as $l_i$,
  the  time length of learner $L_s$ viewing  $V_i$ as $t^s_{i}$.

\begin{figure}\centering
\includegraphics[height=2.2   in,width=6.     in,angle=0]{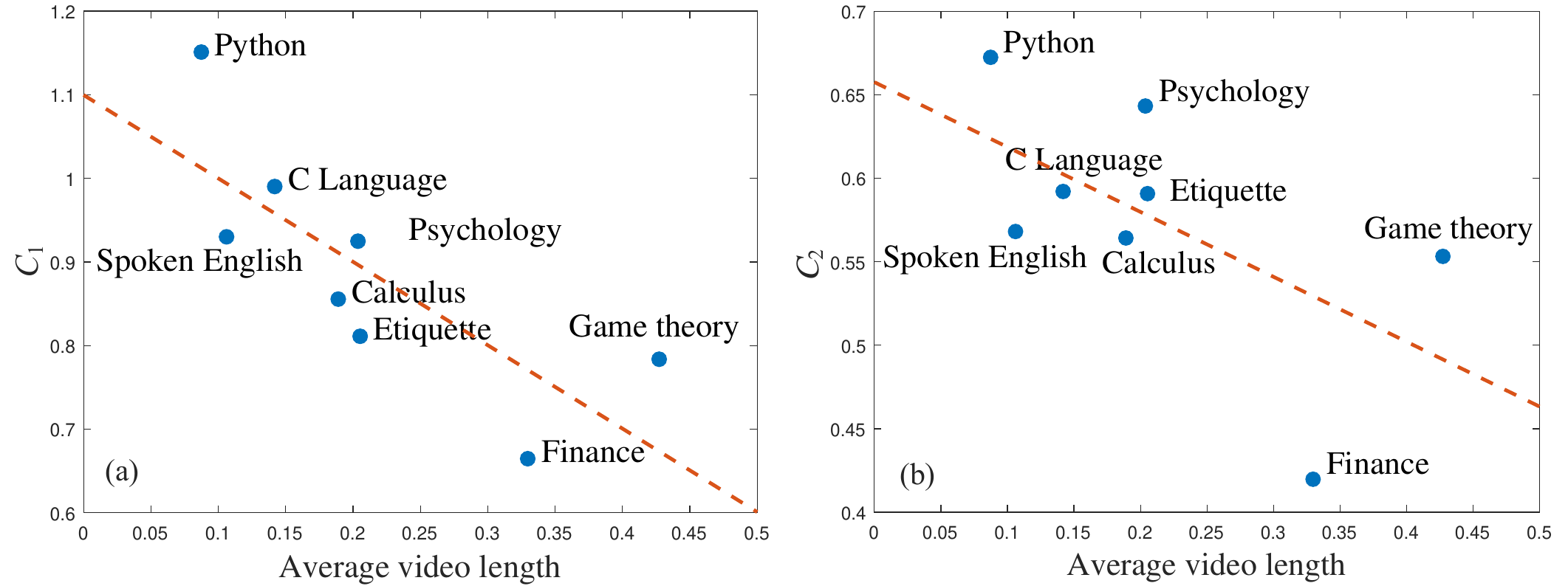}
\caption{  { \bf The negative correlation between $
C_l$ ($l=1,2$) and average video length (unit: hour).
}   Under the assumption  that  learners tend to    view  whole videos,
the  completion rate of viewing videos per learner  can be measured    by   $
C_l$ ($l=1,2$), which are defined in the last paragraph of Section 2.
 } \label{fig1}      
\end{figure}

Calculate the   relative viewing time length  (compared with video lengths) per learner,
 and the
    number of  videos viewed by per learner.
Under the hypothesis that   learners tend to    view    whole videos,
     the ratio between these  two averages  $
C_1=    \left( {\sum_{s=1}^m  \sum^n_{i=1} t^s_{i}/l_i   } \right)/{\sum_{s=1}^m |S^V_s| }
$
      measures  the
completion rate of viewing videos per   learner.
 This rate     can also be measured     by $C_2=1/n \times \sum_{i}  \sum_{s\in S^L_i} \min(t^s_i/l_i,1)/|S^L_i| $  at the same hypothesis.
 Both $C_1$ and $C_2$
  negatively
correlate  to average video length~(Fig.~\ref{fig1}). In fact, human attention spans are limited. A long video is hard to attract learner attentions  from beginning to end.
It means     a long  video's content  should be carefully designed if
its length cannot be shortened.



\section*{Categorization of learners}

 Learners' attention spans related to a course are different. Some are motivated to learn the whole courses, and others part of contents\cite{Wang,Jordan,Alraimi}. Therefore, learners  can be sketchily categorized as two viewing types, namely  segment-learners and
potential all-rounders. Discussing the factors of dropout and engagement for segment-learners has limited insight,
but is meaningful for potential all-rounders. To narrow the scope of finding potential all-rounders, we provided a non-parametric method of categorizing learners  based on  their  viewing time length.

\begin{figure}\centering
\includegraphics[height=3.3 in,width=6.4   in,angle=0]{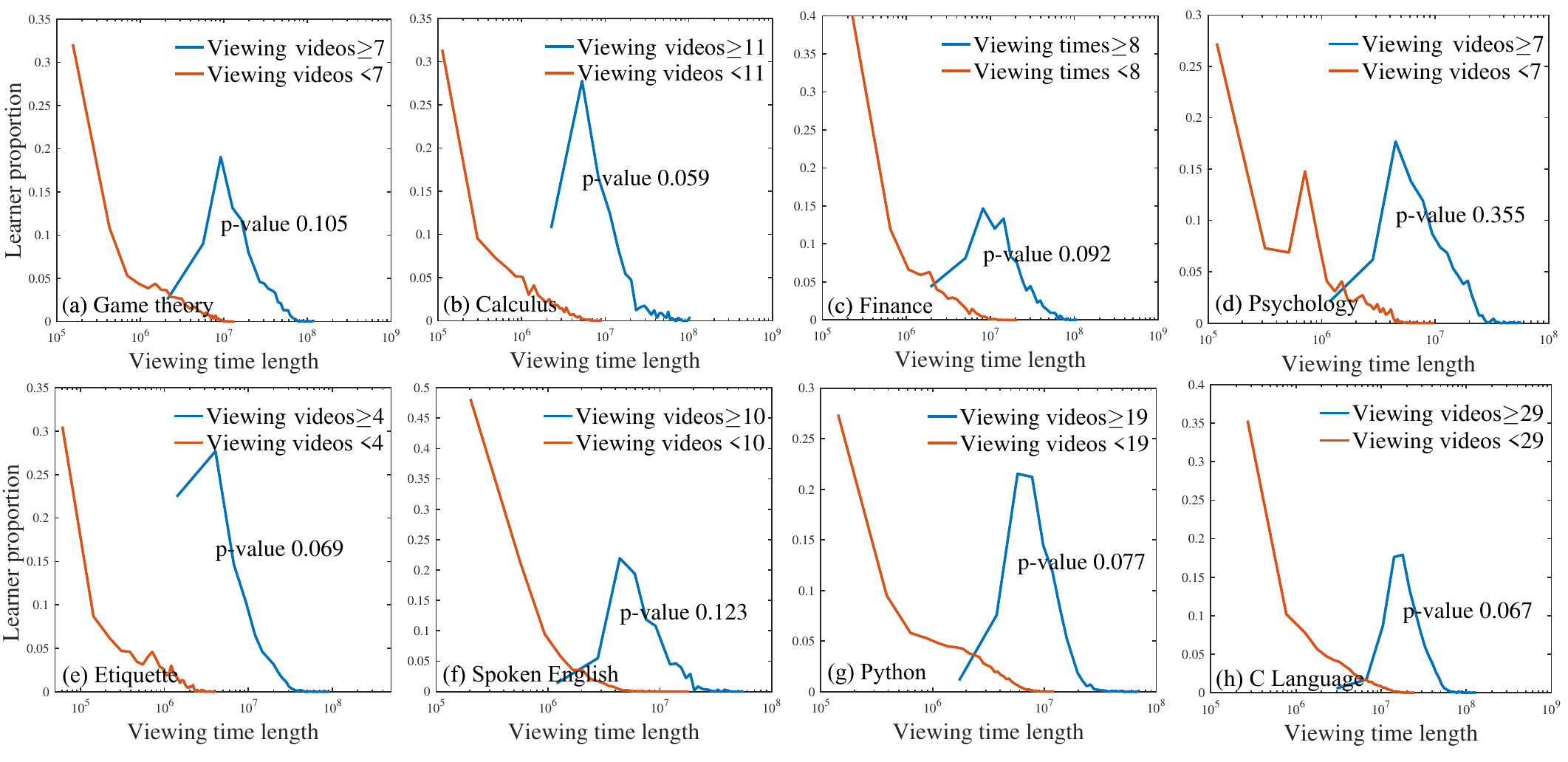}
\caption{  { \bf  The viewing time length distributions of learners. }
Panels showed the distributions of lognormal-learners (blue lines) and those of  other learners (red lines). At significance level 5\%, KS test cannot reject the null hypothesis that the viewing time lengths of lognormal-learners follow a lognormal  (p-values$>0.05$).} \label{fig2}      
\end{figure}

Table~\ref{tab1} showed
 the number of all-rounders is   very small for each course. However, even the learners, who decided to complete a course, might not view all videos. For such a learner, his tenacity   of viewing   videos    could be compared to a unit   whose failure mode is of a fatigue-stress nature. The life of such a unit follows a lognormal distribution\cite{Gaddum}. And the tenacity   of a learner      could be measured by his viewing time length. We labelled  the learners  whose viewing time length follows a lognormal distribution  as lognormal-rounders. In Table~\ref{tab2}, we provided an algorithm to recognize them.
   Fig.~\ref{fig2} showed the results of the algorithm   applied to the empirical data. Specific statistical indexes of the two types of learners were listed in Table~\ref{tab3}.


 \begin{table*}[!ht] \centering \caption{{\bf An  algorithm   of categorizing  learners.} }
\begin{tabular}{l r r r r r r r r r} \hline
Input: the viewing time length  $t_s$ and   the number of viewed videos  $n_s$  of learners $L_s$ $(s=1,...,m)$.\\
\hline
For   $k$ from $0$ to $\max(n_1,...,n_m)$ do: \\
~~~~Do  KS test for $t_s$ of the learners $L_s$ satisfying  $n_s> k$
 with   the null
 hypothesis that \\they follow a lognormal distribution;\\
~~~~Break  if  the test cannot reject  the null hypothesis  at   significance level  $5\%$. \\ \hline
Output:   the current $k$ (denoted as $\kappa$). \\ \hline
 \end{tabular}
   \begin{flushleft}
 The unit of time is
millisecond. If    $n_s>\kappa$  then
   $L_s$ is labelled as a  lognormal-rounder.
   \end{flushleft}
\label{tab2}
\end{table*}


\begin{table*}[!ht] \centering \caption{{\bf Specific statistical indexes of the empirical data. } }
\footnotesize\begin{tabular}{l rrrrrr rr } \hline
 Course &Category  & $a$ &	$b$&	$c$ & $d$ & $e$  \\ \hline
  \multirow{2}{*} {\it Calculus}  &$A$&	569	& 28.120	&3.848 &	4.078	&23.06  \\
& $B$	&2,386&	3.302	&0.319	&1.083	&2.385 \\\hline
  \multirow{2}{*} {\it Game theory}& $A$ &1,522	&16.86&	5.919&	3.578	&14.13\\
& $B$	&3,242	&2.581&	0.510&	0.872 	&1.689\\\hline
 \multirow{2}{*} {\it Finance}&$A$ &	1,057	&21.54&	5.501	&3.820&	16.20\\
&$B$ 	&5,323&	2.157 	&0.478	&0.602	&1.276 \\\hline
 \multirow{2}{*} {\it Psychology} &$A$&799	&15.30	&3.100&	3.531&	13.30\\
	 &$B$ &3,028	&2.294	&0.336&	0.648	&1.643 \\\hline
\multirow{2}{*} {\it Spoken English} &$A$&583	&19.61	&2.426 	&3.724 &	18.45&\\
  &$B$	&11,136&	2.164 &	0.211&	0.636&	1.670\\\hline
\multirow{2}{*} {\it Etiquette} &$A$&2,084	&12.95&	2.213 	&3.084	&10.80\\
&$B$&	1,762	&1.683&	0.157&	0.469	&1.035\\\hline
\multirow{2}{*} {\it C Language} &$A$ &2,367	&46.57 	&6.609 	&5.161 	&43.24\\
&$B$&	15,174	&7.147&	0.750 &	1.827	&5.833 \\\hline
\multirow{2}{*} {\it Python} &$A$&2,549&	28.76	&2.748 &	4.475 	&28.75\\
&$B$&	10,868	&5.600&	0.461	&1.791 &	5.243 \\
\hline
 \end{tabular}
  \begin{flushleft}
  $A$: lognormal-learners,   $B$:  other learners,   $a$: the number of learners,   $b$:
the number of videos viewed by per learner,   $c$: the viewing time length per learner (unit: hour),   $d$: the  entropy per learner,
 and $e$: the
geometric mean~(\ref{eq2}) per learner.\end{flushleft}
\label{tab3}
\end{table*}

%



 \section*{MOOC attraction  measurements}



When a learner views a course, we can regard the video he chooses to view   as a random event, and so  the label of the chosen video as a random variable. When the order of course contents is ignored,
 the more videos a learner views, the more even his viewing time distributes, then the higher  the uncertainty of which video is viewed in a viewing event is.
  Entropy can be used to  measure the uncertainty\cite{Shannon}.
 Denote $X_s$ to be the label of the video  chosen by a viewing event of learner $L_s$.
The probability of
  choosing
 video $V_i$ is $p(X_s=i)=t^s_i/\sum^n_{j=1} t^s_j$, and so the
 entropy of $X_s$ is
\begin{align}\label{eq1}  H(X_s)&=-\sum^n_{i=1}p(X_s=i)\log_2 p(X_s=i) .
\end{align}
We can see that
if  $L_s$ views a new video in a short time, then $H(X_s)$   increases a little.
Therefore, the  number of  videos viewed by $L_s$ can be measured by  $2^{H(X_s)}$
  in a continuous way, which overcomes the shortcoming brought by the discreteness of counting viewed videos.


The
entropy is free of the viewing time length  $\sum^n_{j=1} t^s_j$. However, the attraction of a course to a learner often positively correlates to the time he spent on the course. We should integrate his  entropy and viewing time length into one index to measure the attraction to him. If the lengths of all videos are equal, the unit of  $2^{H (X_s)}$ and that of the relative viewing time length $ \sum^n_{i=1}  {t^s_i}/{l_i}$ are the same, namely the length of one video. Hence we can use their  geometric mean as an index of measuring course attraction:
 \begin{align}\label{eq2}  I(X_s)&= \left( 2^{H(X_s)}  \sum^n_{i=1} \frac{t^s_i}{l_i} \right)^{\frac{1}{2}} .
\end{align}
The reasonability of the formula~(\ref{eq2}) could be illustrated through following examples.

  Learner $L_s$ viewed   video $V_1$ with time length $t_1=l_1$, then his entropy $H(X_s)=0$, and geometric mean $I(X_s)=1$.
If he viewed    $V_1$ and $V_1$ with time length $t_1=l_1$, $t_2=l_2$, then   $H(X_s)=1$,  $I(X_s)=2$.
If   $t_1=2l_1$, $t_2=l_2$, then   $H(X_s)=0.92$,  $I(X_s)=2.38$.
If he viewed $V_i$, $i=1,2,3$  with time length $t_1=l_1$, $t_2=l_2$,  $t_3=l_3$ then   $H(X_s)=1.59$,   $I(X_s)=3$.
As above schematic examples showed, the geometric mean  $I(X_s)$ profiles the diminishing marginal utility
in learning, because  $\partial^2 I(X_s)/\partial (t^s_i)^2<0$.
Formula~(\ref{eq2})  also profiles the increasing process of information in the process of  viewing new videos,
  because  $(p_1+p_2)\log(p_1+p_2)- (p_1\log p_1 +p_2 \log p_2)>0$.

\begin{figure}\centering
\includegraphics[height=1.7 in,width=6.4   in,angle=0]{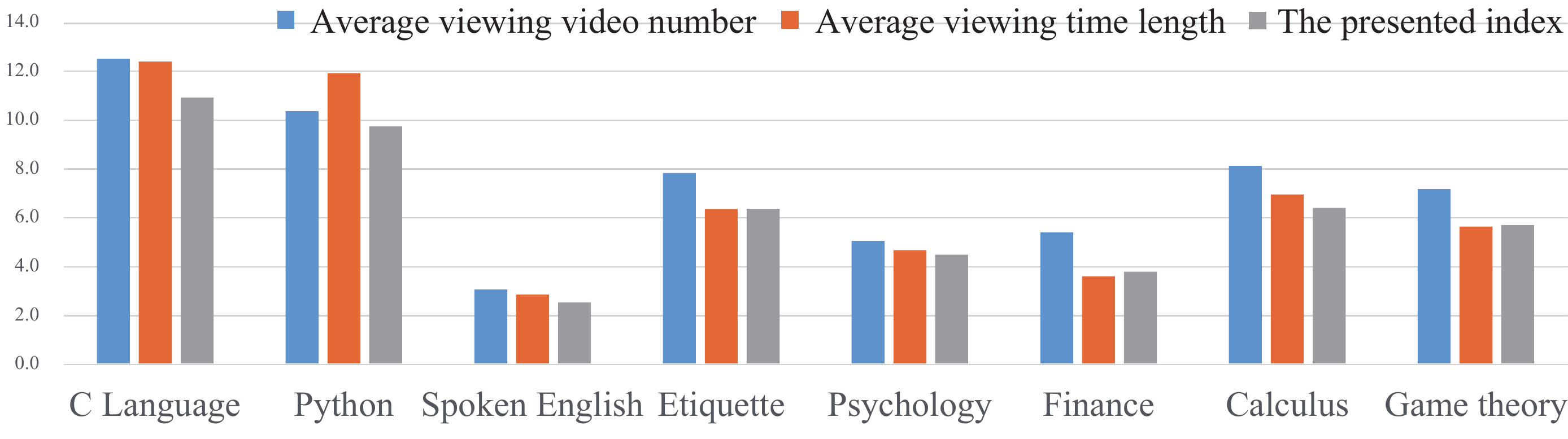}
\caption{  { \bf Course attraction indexes.
} For each course, the first two averages are calculated over all learners, and the presented index is average   geometric mean (\ref{eq2}) of all learners   $\sum^m_{s=1}I(X_s)/m$.
 } \label{fig3}      
\end{figure}

The eight courses were selected from different fields. Some popular courses, such as {\it Python,
Spoken English}, can attract numerous learners. Meanwhile, some theoretical courses, e. g.~{\it Calculus},
can hardly attract  the learners without corresponding prior knowledge. Hence, to compare attractions
of courses from different fields, it is suitable to use the average of the geometric means over all learners  $\sum^m_{s=1}I(X_s)/m$,  which removes the heterogeneity of course learner numbers.
    Fig.~\ref{fig3} showed that this average   positively correlated to the number of videos viewed by per learner and to the viewing time length per learner, which fits the common sense: view more and longer, be attracted deeper.

\begin{figure}\centering
\includegraphics[height=3.   in,width=6.4   in,angle=0]{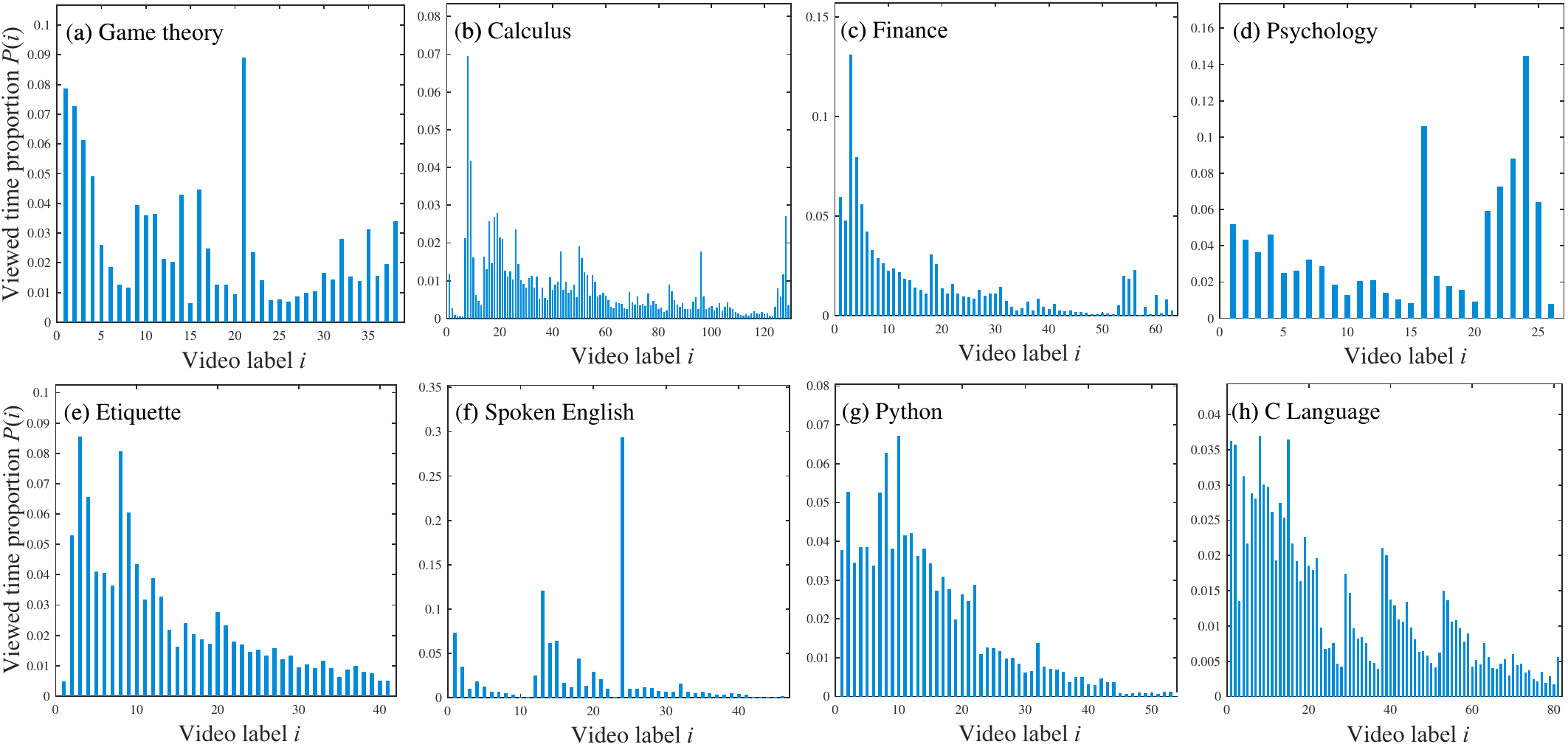}
\caption{  { \bf  Viewing time distributions on videos.
}The  entropy   calculated based on the  viewing time distribution of a  course   describes  the balance of  the course's attraction  over videos.
 The more uniformly the viewing time of a course distributed, the larger the entropy is.
 } \label{fig4}      
\end{figure}

\begin{figure}\centering
\includegraphics[height=1.6   in,width=6.4    in,angle=0]{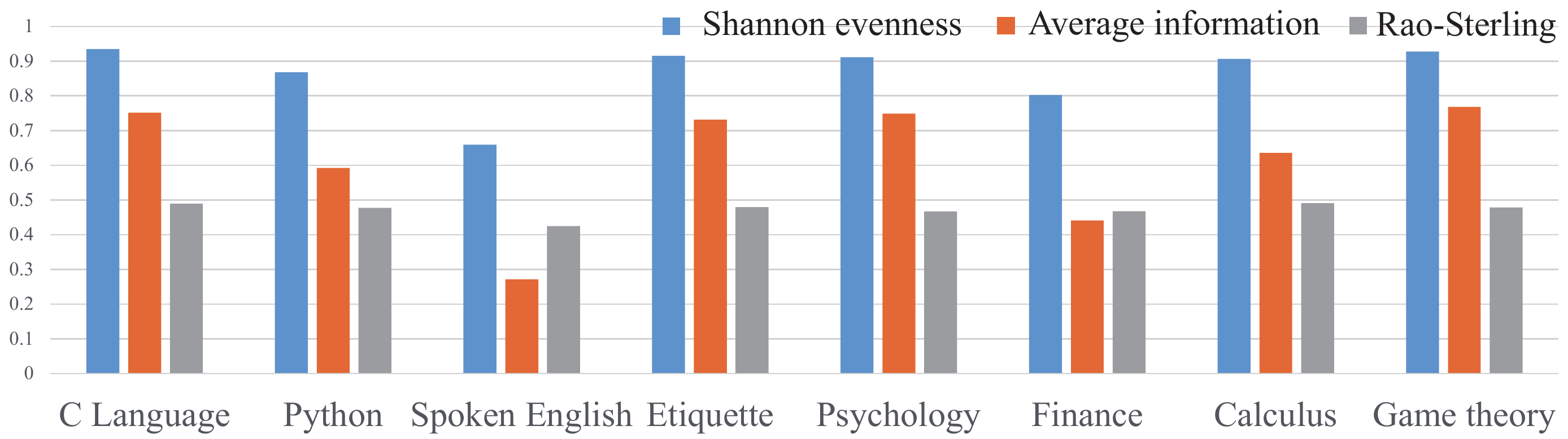}
\caption{  { \bf Balance indexes of course attractions.
}  For each course, those indexes are calculated based on the distribution of its learner viewing time.
 } \label{fig5}      
\end{figure}

Now let us  discuss the balance of a course's attraction  over videos.
Consider  a course with videos $\{V_1,V_2,...,V_n\}$,  and denote $l_i$ to be the length of
   viewing time   spent by course learners  on $V_i$ ($i=1,...,n$).
Then the entropy $H =-\sum^n_{i=1} P(i)\log P(i)$
  profiles the attraction  balance of the course~(Fig.~\ref{fig4}), where  $P(i)=l_i/\sum^n_{j=1}l_j$.
 However, courses could have different video numbers.
 Suppose  two courses' viewing time are all   distributed uniformly on videos.
Then  the entropy of the course with more videos is larger than that of the course with fewer videos.
Therefore, to compare the attraction  balances of courses, we should remove the
 heterogeneity of  the video numbers of courses, which can be achieved  by Shannon evenness $H /\log_2 n $\cite{Pielou}, or  by
  $2^{H }/n$~(Fig.~\ref{fig5}).

The two indexes of balance remove the heterogeneity of learner numbers and that of video numbers.
These indexes of {\it Spoken English} were relatively low, which is due to that 30\% viewing time was attracted by one video (Fig.~\ref{fig3}(f)).
 For a course, low indexes of balance imply that the course cannot attract learners persistently,
 and so its teachers could improve the contents of less viewed videos.
 Rao-Sting operator \cite{Stirling}
 $\Delta=\sum_{i,j(i\neq j)} d^\alpha_{ij}  P(i)^\beta P(j)^\beta$ (where $\alpha=\beta=d_{ij}=1$ for all possible  $i$ and $j$)
can also portray the balance of a viewing time distribution, but it does not take into account the difference of course video numbers.

  \section*{Correlation between teaching order and viewing order}

Designing teaching order is fundamental in pedagogy. MOOC education is learner-driven rather than knowledge-oriented. The correlation between teaching order and learning order affects course quality. Learning order can be reflected by viewing order, especially when learners are unsupervised. Teaching order can be expressed by video labels. If nearly viewed videos have close labels, the learning order is consistent with the teaching order. Surely, the order correlation is not meaningful to some humanities courses such as {\it Spoken English}, but is important to some natural science courses such as  {\it Calculus}.

\begin{figure}\centering
\includegraphics[height=1.7   in,width=6.4   in,angle=0]{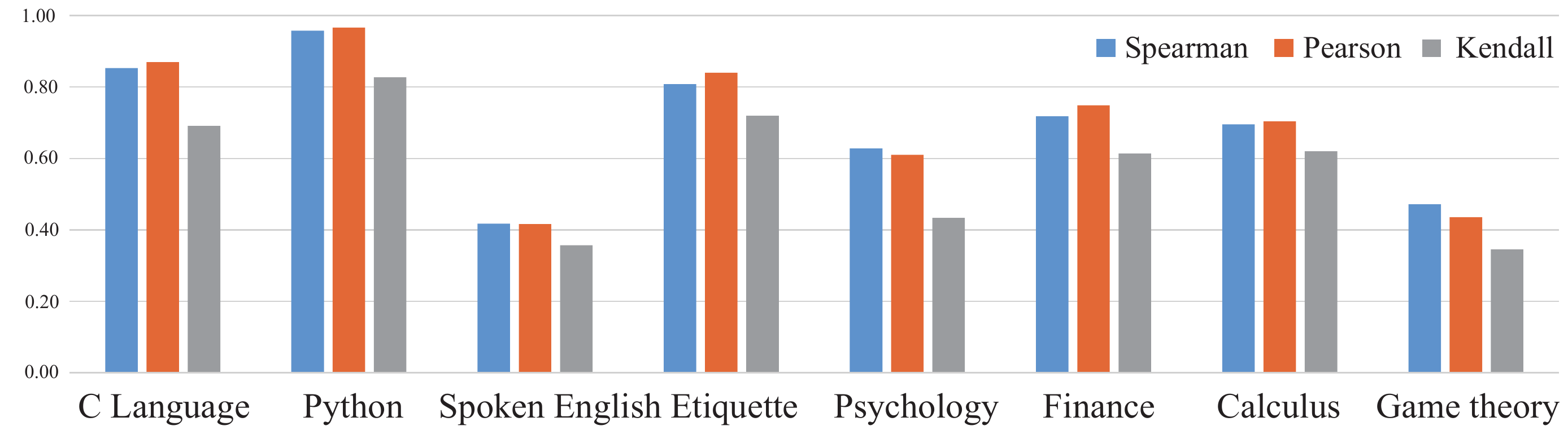}
\caption{  { \bf   Correlation between teaching order and learning order.
}  Teaching order is expressed through video labels, and learning order is expressed by viewing order. The correlation is measured by three typical correlation coefficients between variable pair $i$ and $\nu(i)$ (Eq.~\ref{eq3}).
 } \label{fig6}      
\end{figure}

We provided a method to measure   the   order correlation.
For each leaner $L_s$, we   measured the viewing correlation between any  two videos  $V_i$ and $V_j\in S^V_s$ (the set of videos  he viewed) through
  $w^s_{ij}=f(|\tau^s_i-\tau^s_j|)$, where $f(\cdot)$ is a nonnegative and decreasing function, $\tau^s_i$ and $\tau^s_j$ are the start times
of   $L_s$   viewing $V_i$ and $V_j$ respectively.
 A small  value of $|\tau^s_i-\tau^s_j|$ implies  it is   likely to exist a viewing order between  $V_i$  and   $V_j$.
For each video $V_i$, we calculated the   weighted summation  \begin{align}\label{eq3} \nu(i)=\frac{\sum_{s\in S^L_i}   \sum_{j\in S^V_s\backslash i  }w^s_{ij}j  }{\sum_{s\in S^L_i}    \sum_{j\in S^V_s\backslash i  }w^s_{ij} }.  \end{align}
 The correlation coefficient between video label  and
the   weighted summation~(\ref{eq3})  measures
the  correlation between teaching order and viewing order. Here we let
  $w^s_{ij}= \min \left(  {24}/{|\tau^s_i-\tau^s_j|}, 1   \right)$, and calculated   three widely used correlation coefficients\cite{Kendall, Hollander}
for the eight courses. Fig.~\ref{fig6} showed that the three correlation coefficients of {\it Spoken English} were relatively low, which is consistent with common senses. However,
 if these correlation coefficients of a mathematic course are low, then  the teaching order of the course   needs to be redesigned.

\begin{figure}\centering
\includegraphics[height=3.   in,width=6.4   in,angle=0]{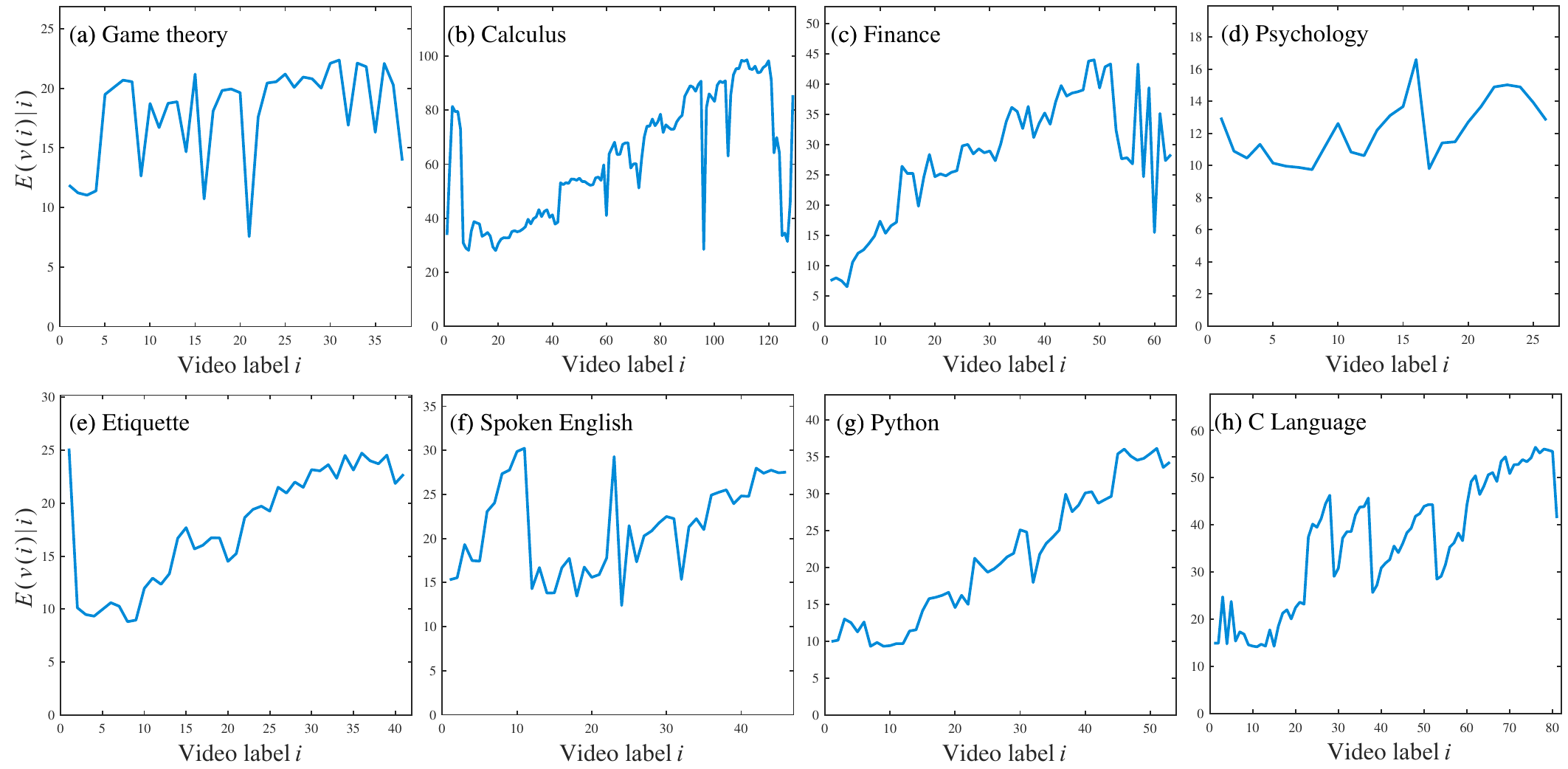}
\caption{  { \bf  The
conditional expected value of $\nu(i)$ given video label $i$.
} The approximatively linear trend of $E(\nu(i)|i)$ guaranteed the effectiveness of correlation coefficients   in Fig.~\ref{fig5}.
 } \label{fig7}      
\end{figure}

Note that the Pearson coefficient
indicates the strength of a linear
relationship between two variables $X$ and $Y$, unless   the conditional expected value of $Y$
given $X$ (denoted  as $E(Y|X)$) is   linear or approximate linear
in $X$,  and verse vice.
The visual examinations shown in Fig.~\ref{fig7}     guaranteed     the effectiveness of correlation analysis addressed here.

\section*{Discussion   and conclusions}


 MOOCs are examples of learner-centered and autonomous learning.
 MOOC learning behaviors tend to be individual, unsupervised, and nonintervened. Human behaviors in such situations often reveal what they are. Therefore,
one could have faith in the reliability   of viewing behavior data.
 We employed viewing behavior data to categorize learners, to assess course attractions and the correlations between teaching order and learning order. The practicability of our methods is validated with  the empirical data provided by iCourse.
 Our results help to understand the rules of human  cognition behaviors   on the Internet.


Our methods need further improvement. In terms of data application, learning preferences should be addressed in learning pattern recognition, which helps teachers to implement individualized education. In terms of pedagogy, studying autonomous and learner-centered MOOC learning helps the development of constructivism theory\cite{Tobias,Huang2002} and provides cases for online pedagogy. In terms of data fusion, testing and certificating behaviors should be considered in MOOC profile, which contribute to assess learning achievements. The analysis of correlations between these behaviors and viewing behaviors helps to inference the achievements of the learners without certifications and test scores.

 We finished our   study by asking a question: How to assess MOOCs.
 The indexes to assess traditional courses are inappropriate for profiling MOOC quality such as course completion rate\cite{Sandeen}. It is, therefore, necessary to design new indexes to assess MOOCs in their own way.
 Assessing a course is in essence to determine the degree to which its teaching reaches its goal\cite{Tyler}, and so inextricably connects with learning   quality of its learners\cite{Allen}, learner engagements\cite{Henrie,Hew2016}, learning patterns\cite{LiLY,Kizilcec1,Evans}, and achievements\cite{Cheng,Coetzee}. Therefore,
learning achievements  contributes to MOOC quality, and so our results  have
potential to be      indexes of  MOOC assessments.

\section*{Funding}
ZX acknowledges support from      National   Science Foundation of China (NSFC) Grant No. 61773020.

\end{document}